\documentclass[a4paper,11pt]{article}
\usepackage{pos}
\usepackage{caption}
\usepackage{subcaption}

\title{Testing the AGN Radio and Neutrino correlation using the MOJAVE catalog and 10 years of IceCube Data}
 \ShortTitle{MOJAVE AGN radio neutrino correlation}

\author{The IceCube Collaboration \\{\normalsize \normalfont(a complete list of authors can be found at the end of the proceedings)}}

\emailAdd{adesai@icecube.wisc.edu}

\abstract{On 22 September 2017, IceCube reported a high-energy neutrino event which was found to be coincident with a flaring blazar, TXS 0506+056. This multi-messenger observation hinted at blazars contributing to the observed high-energy astrophysical neutrinos and raised a need for extensive correlation studies. Recent work shows that the internal absorption of gamma rays, and their interactions intrinsic to the source and with the extragalactic background, will cause a lack of energetic gamma-ray and neutrino correlation while hinting towards a correlation between neutrinos and lower photon energy observations in the X-ray and radio bands. Studies based on published IceCube alerts and radio observations report a possible radio-neutrino correlation in both gamma-ray bright and gamma-ray dim active galactic nuclei (AGN). However, they have marginal statistical significance due to limited available data. We present a correlation analysis between 15 GHz radio observations of AGN reported in the MOJAVE XV catalog and 10 years of IceCube detector data and discuss the results derived from a time-averaged stacking analysis.

\vspace{4mm}
{\bfseries Corresponding authors:}
Abhishek Desai$^{1*}$, Justin Vandenbroucke$^{1}$, Alex Pizzuto$^{1}$\\
{$^{1}$ \itshape University of Wisconsin Madison}\\[4mm]
$^*$ Presenter

\FullConference{37$^{\rm{th}}$ International Cosmic Ray Conference (ICRC 2021)\\
		July 12th -- 23rd, 2021\\
		Online -- Berlin, Germany}

}

\begin{document}
\maketitle

\section{Introduction}\label{sec:intro}

The IceCube Neutrino Observatory is a cubic kilometer neutrino detector installed in the ice at the geographic South Pole \cite{icecube_2017_intro}. As neutrinos arrive at Earth, they may interact with the ice or surrounding bedrock leading to the creation of secondary charged particles. These particles produce Cherenkov radiation that can be detected by the digital optical modules connected to strings in the ice and used for reconstruction to derive the direction, energy, and flavor of the neutrinos. In 2013, IceCube reported the detection of highly energetic neutrinos of astrophysical origin \cite{icecube_2013_astrophysical_neutrino}. While the nature of the sources producing astrophysical neutrinos is yet not known, one of the prime candidates suspected in producing such neutrinos are Active Galactic Nuclei (AGN). In 2017, IceCube detected a neutrino event originating from the direction of an AGN, TXS 0506+056, which was found to be flaring in the gamma ray regime at the time of the event. Multiple follow-up analyses have been launched in order to help us understand if AGN are responsible for the production of these astrophysical neutrinos, and how they are produced. Multi-messenger analyses like \cite{TXS_Icecube,Plavin_2020a,Plavin_2020b,OVRO_neutrinos_2020,zhou2021neutrino_agn} which study the correlation of IceCube neutrinos with the observed electromagnetic radiation at different wavelengths from AGN help us understand the processes that could give rise to neutrinos in AGN. 

One of the theories that could predict the production of high-energy neutrinos in AGN is discussed recently by an analysis reported by \cite{Plavin_2020a,Plavin_2020b} which looks into a correlation between the seven-year public IceCube data and Very-Long-Baseline radio Interferometry (VLBI) selected AGN. Production of high-energy neutrinos can occur via , hadronic (nucleon-nucleon) or photohadronic (nucleon-photon) interactions in AGN \cite{Eichler78,Plavin_2020a}. In the bright central parsecs of AGN, observable in the radio band, $pp$ interactions are generally suppressed with respect to $p\gamma$ interactions \cite{pp_suppression_1987_sikora,Inoue_2019}. In \cite{Plavin_2020b} it is argued that, if $p\gamma$ interactions are the cause of high-energy neutrinos then the gamma rays, produced in these photohadronic processes alongside with neutrinos, will interact with the same target photons to produce electron-positron pairs which will lead to a chain of processes resulting in gamma rays without enough energy to undergo pair production. This theory, which is also discussed by \cite{OVRO_neutrinos_2020},  can serve as a possible explanation for the lack of high energy gamma-ray (as observed by {\it Fermi}-LAT) and neutrino correlation in AGN and may hint towards a correlation with lower energy gamma-rays in the keV to GeV regime.

Although the theories proposed by analyses like \cite{Plavin_2020a,Plavin_2020b,OVRO_neutrinos_2020} are very promising, they are based on hints of radio-neutrino correlation with marginal statistical significance due to the limited data available. It makes it necessary to follow up with enough statistical power to test the authenticity of this correlation.  For a source with an energy spectrum of the same slope as the measured diffuse muon neutrino spectrum ($\approx E^{-2.2}$), there are two orders of magnitude more astrophysical neutrinos in the complete IceCube data-set than in the public high-energy alert event sample. Using the full IceCube data thus increases both neutrino signal and background data thereby improving the statistics of the study and dealing with the limitations seen in the previous analysis, making it important to perform a follow-up analysis. 

In this work, we make use of the data provided by the MOJAVE XV catalog \cite{MOJAVEXV} to search for correlations with ten years of IceCube data. This catalog dataset has observations of AGN with varying cadence allowing us to perform two separate studies on it (time-integrated and time-dependent). While the time-integrated analysis makes use of the average flux density values of the MOJAVE sources, the time-dependent case will make use of a time binned lightcurve, derived from the MOJAVE observations, to search for neutrino correlations. The time-dependent study is beyond the scope of this article and will be discussed in a separate work soon. The time-integrated study and its results are described in this article. We also list our results alongside the recent work performed by \cite{Plavin_2020a,Plavin_2020b,zhou2021neutrino_agn} to help us get a better understanding of the studies performed on AGN and the differences seen in the results based on the methodology used for the study.

\section{AGN Source Sample}\label{sec:source_sample}

\begin{figure}
\centering
\begin{subfigure}{0.5\textwidth}
  \centering
  \includegraphics[width=1.1\textwidth]{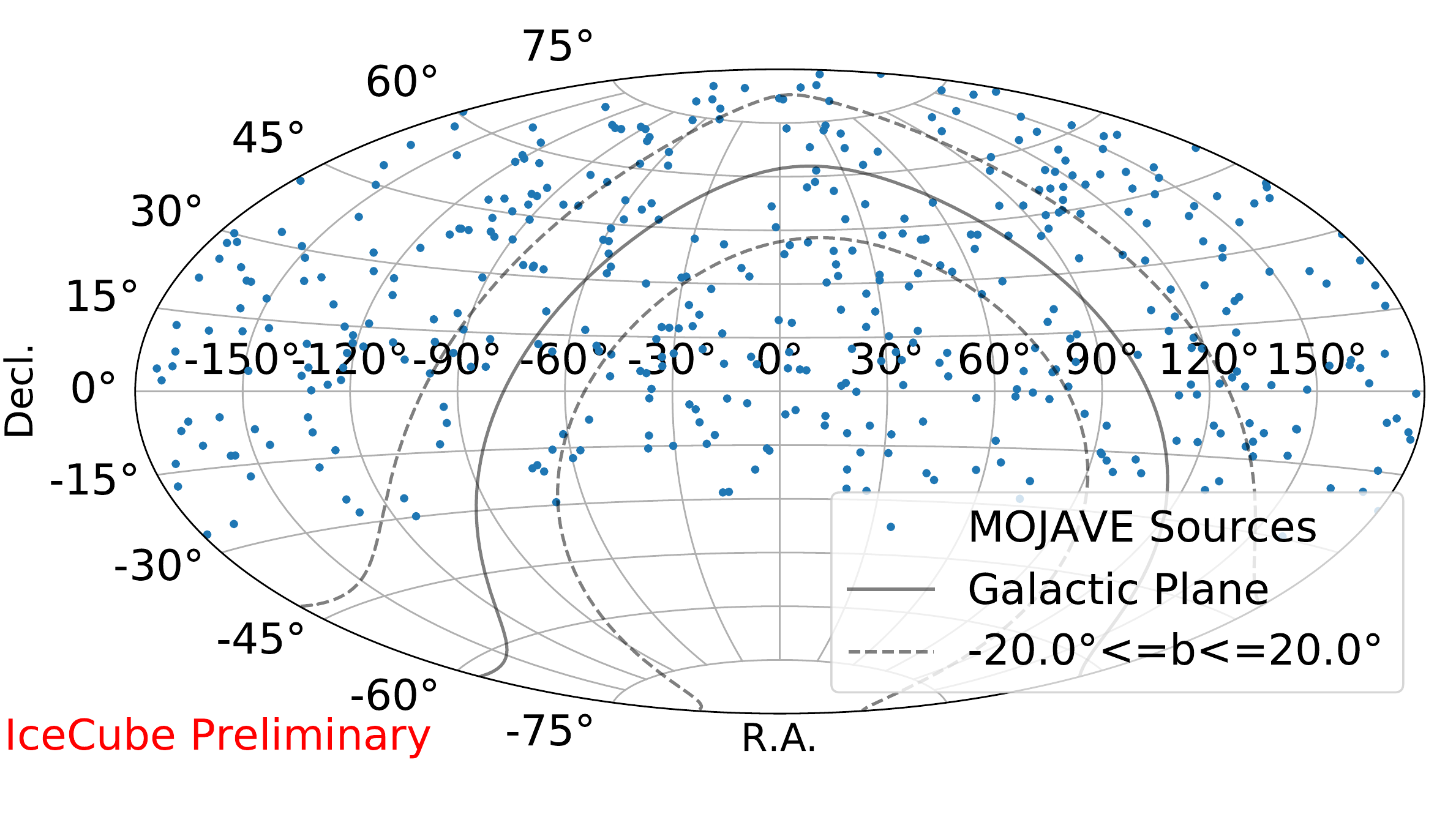}
\end{subfigure}%
\begin{subfigure}{0.5\textwidth}
  \centering
  \includegraphics[width=0.8\linewidth ]{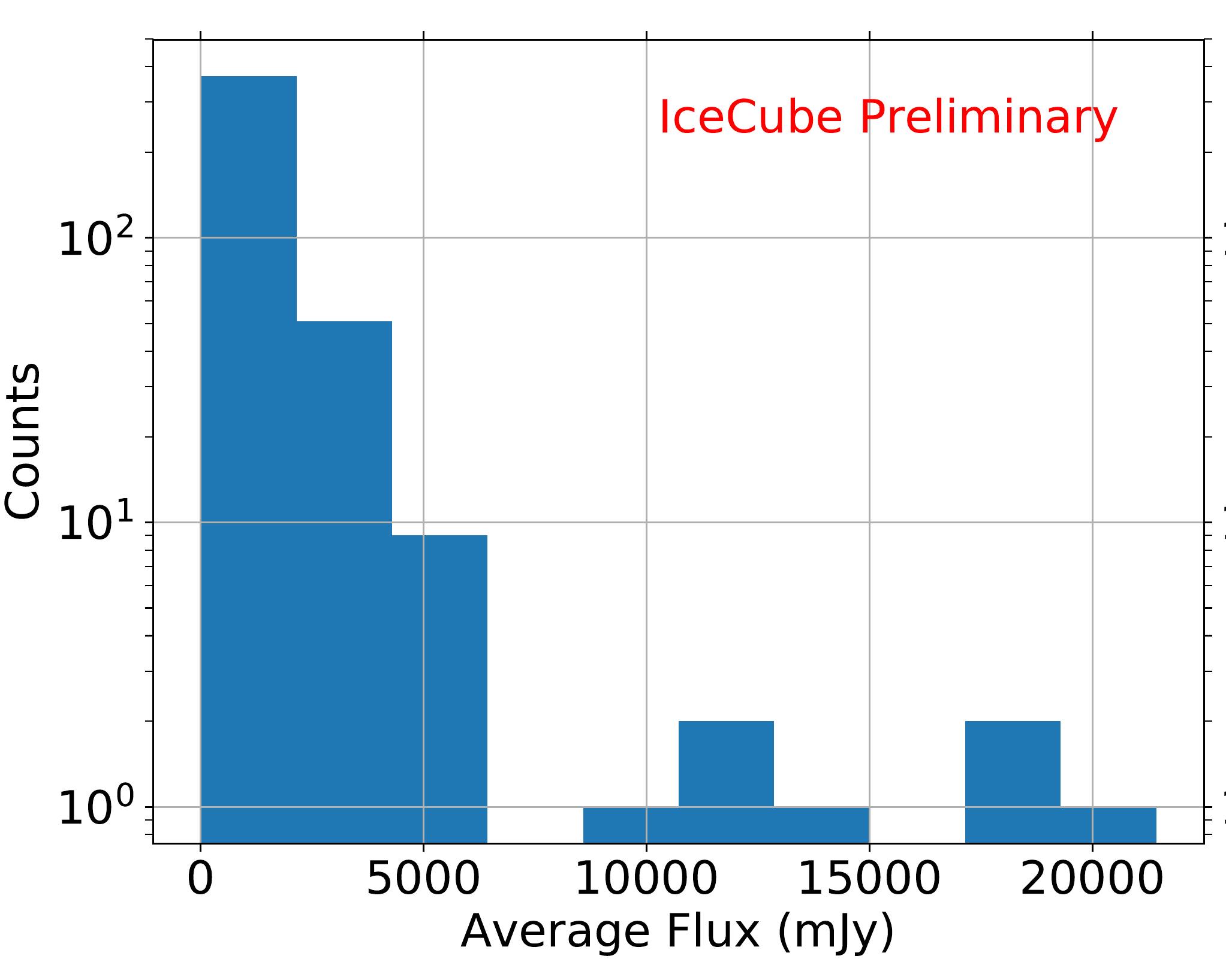}
\end{subfigure}
\caption{Left: Source distribution in Equatorial (J2000) coordinates of the MOJAVE XV dataset consisting of a total of 437 AGNs observed in the 15 GHz band. Note that $\sim75\% $ of the sources lie at positive declination while the rest of the sources lie between $0$\textdegree \,to $-30$\textdegree \,declination. IceCube is more sensitive at the horizon and in the Northern Celestial Hemisphere. Right: Average flux histogram of the 437 AGN observed at 15 GHz which is used as the weight for the stacking analysis.\vspace{-4.5mm}}
\label{fig:aitoff}
\end{figure}

The AGN radio catalog used for this analysis is the MOJAVE XV catalog \cite{MOJAVEXV}, which consists of 5321 observations of 437 AGN in the 15 GHz band using the Very Long Baseline Array (VLBA) in full polarization between 1996 January 19 and 2016 December 26 (sky distribution shown in Fig.~\ref{fig:aitoff}). 
Out of the 437 AGN presented in the catalog, 392 sources are blazars while the rest of the sources are radio galaxies (27 sources), narrow line Seyfert 1  (5 sources) and unidentified AGN  (13 sources). All these AGN have bright compact radio emission with flux densities greater than 50\,mJy at 15\,GHz. While the blazars in the sample are included due to their strong jets, the radio galaxies are included due to a lower redshift value or because of GHz-peaked radio spectra (see \cite{MOJAVEXV} for more details).

\begin{figure}
\centering
\begin{subfigure}{0.5\textwidth}
  \centering
  \includegraphics[width=0.9\linewidth]{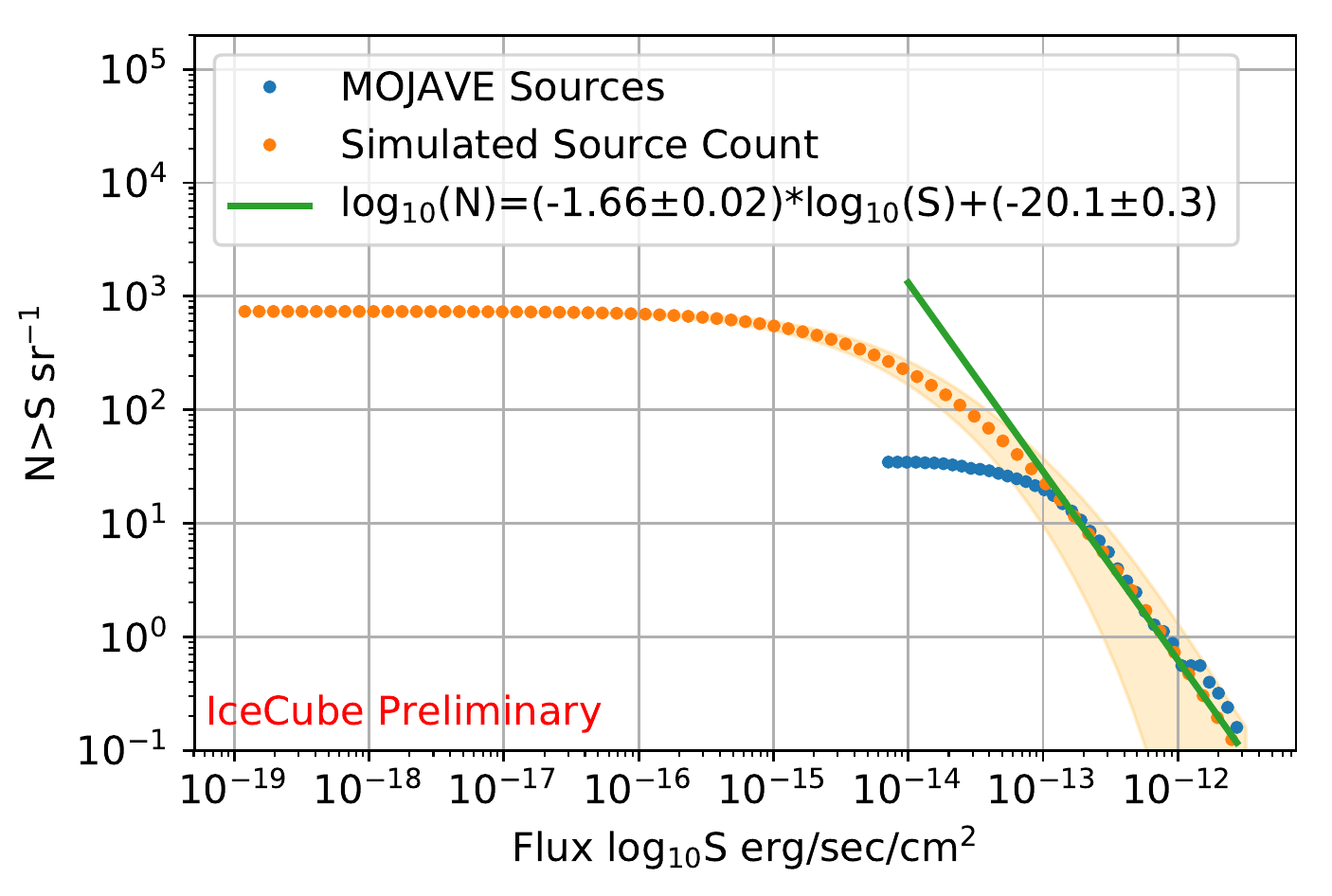}
\end{subfigure}%
\begin{subfigure}{0.5\textwidth}
  \centering
  \includegraphics[width=1.1\linewidth,trim={1cm 3cm 1cm 3cm},clip ]{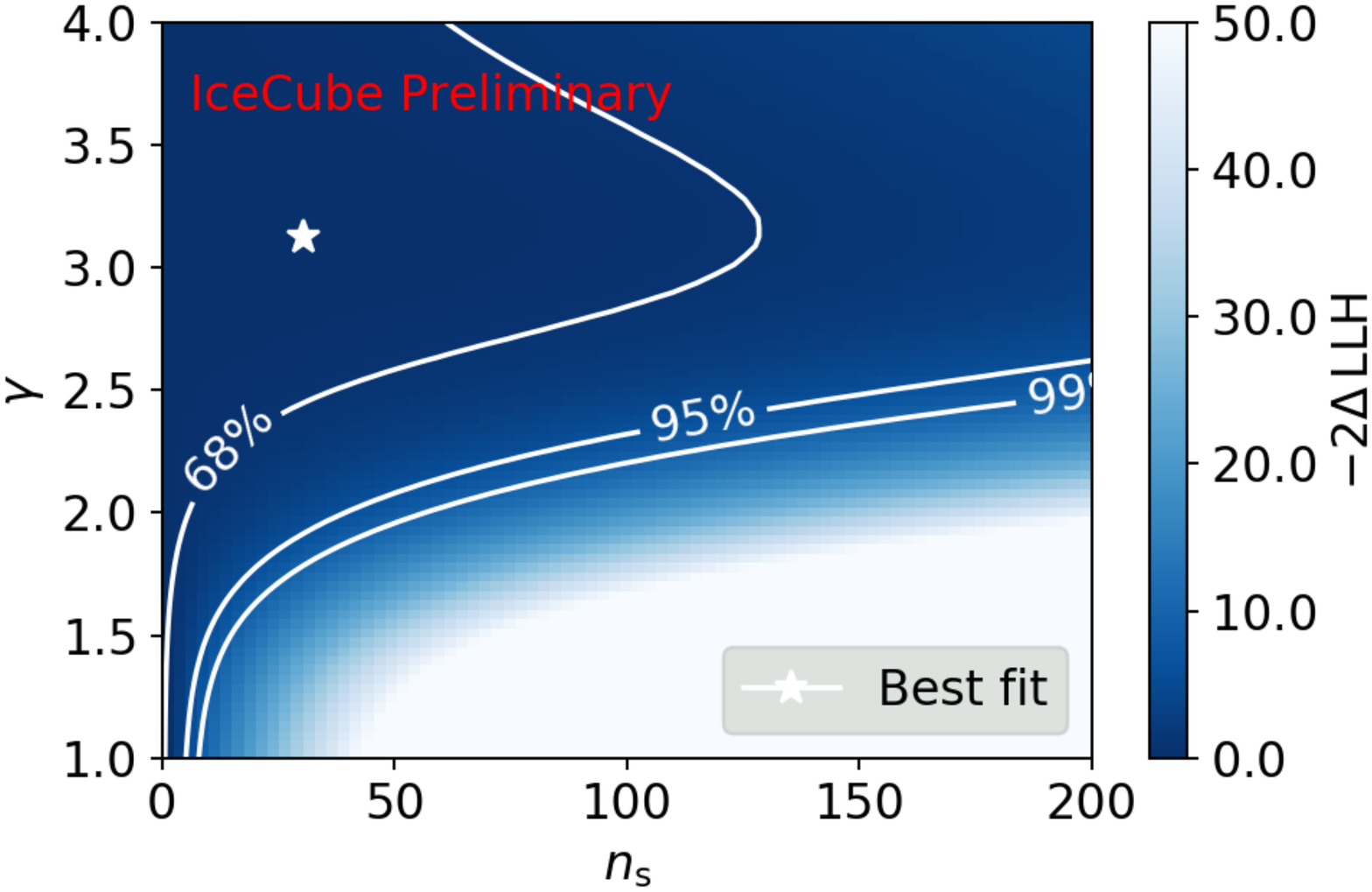}
\end{subfigure}
\caption{Left: Simulated source count distribution of the blazar dominated sample (orange data points) as compared to the source count distribution of the MOJAVE XV sources (blue data points). The shaded region shows the one sigma error to the distribution due to varying the Lorentz factor and viewing angle parameters of the jets. The green line shows the fit at flux densities higher than $10^{-13}$erg sec$^{-1}$cm$^{-2}$ to the simulated sample. Right: The plot shows the contours of the above $n_{s}-\gamma$ likelihood plot for the 1,2 and 3 sigma values from the best fit assuming Wilk's with 2 degrees of freedom. We get a p-value of 0.49 as our result. So, we set an upper limit on neutrino flux from these sources.\vspace{-4.5mm}}
\label{fig:2dlikelihood}
\end{figure}

While the MOJAVE source sample is considered to be complete in terms of VLBA sources observed with high flux densities (>1.5\,Jy) at 15\,GHz, for a larger, unbiased study such as this, it is considered to be a flux-limited sample. Moreover, sources in the MOJAVE catalog are only located at declinations greater than $-30$\textdegree, meaning that a completeness correction is required to account for the spatial limitations and flux limitations of the sample. To estimate the completeness, a source count distribution of the sample is derived by using the radio luminosity function given by the MOJAVE-XVII work \cite{MOJAVEXVII}. The luminosity function is derived from 15\,GHz data of 409 radio-loud AGN observed using VLBA. There is a decrease in the number of sources used by the MOJAVE XVII sample because of an additional condition of a minimum flux density of 0.1\,Jy at 15\,GHz and with at least 5 VLBA observation epochs spaced in time. The Lorentz factor and viewing angle distributions of the source sample (Fig.11 of \cite{MOJAVEXVII}) are used to simulate the sky for any one source class. This is repeated multiple times and the average is taken to derive the source count shown in Fig.~\ref{fig:2dlikelihood} (left). The ratio of the area under the two source count distributions- MOJAVE sources (blue points) and simulated population (orange points)-, is used to derive the completeness of the sample. This is again repeated multiple times to get the average completeness value with an uncertainty value. As the jet properties derived by the MOJAVE XVII study have the same characteristics as the jets observed by the MOJAVE XV sample, the estimated source count distribution can be used to correct the flux-limited sample. Moreover, the source count distribution is derived per steradian $(1/sr)$, ( Fig.~\ref{fig:2dlikelihood}; left) accounting for the spatial limitations of the sample. However, the MOJAVE catalog is a blazar dominated sample, which would mean that the completeness correction will also lead to a sample which is blazar dominated.

\section{Likelihood analysis using {IceCube} data:}\label{sec:likelihood}

A time-integrated stacking analysis is performed using a method similar to the unbinned likelihood-ratio method of \cite{Braun_2008}. The neutrino  track-like event data can be modeled by two main hypotheses: $H_0$: background atmospheric events from atmospheric neutrinos and atmospheric muons, and $H_S$: background atmospheric events in addition to signal events from astrophysical neutrino sources. Using the known spatial and energy distributions of background and astrophysical neutrinos, the probability density functions (PDFs) $P({\rm Data}|H_0)$ and $P({\rm Data}|H_S)$ are calculated. The test statistic is defined as the log of the likelihood ratio  between the null hypothesis and the best-fit alternative hypothesis as ${\rm TS} = 2 \log\left[{P({\rm Data}|H_S)}/{P({\rm Data}|H_0)} \right]$.
Given a candidate astrophysical source at location $\vec{x_s}$, we use the TS to test its incompatibility with the null hypothesis $H_0$. 
This is done by using a set of $N$ neutrino data events, each with a reconstructed direction $\vec{x_i}$ and reconstructed energy $E_i$. Each of these is assigned a probability of the event belonging to the source being analyzed. Assuming a source neutrino spectrum given by a power-law energy spectrum $E^{-\gamma}$, the source PDF is given by 
\begin{equation}
    \mathcal{S}_i(\vec{x_i},\vec{x_s},E_i,\gamma) = \frac{1}{2\pi\sigma_i^2}\exp\left({-\frac{|\vec{x_i}-\vec{x_S}|^2}{2\sigma_i^2}}\right) P(E_i|\gamma)\, ,
\end{equation}
where $\sigma_i$ is the angular reconstruction error estimate and $P(E_i|\gamma)$ is the probability of observing a reconstructed muon energy $E_i$ given a source spectral index of $\gamma$ (see \cite{Braun_2008}). Note that the source position uncertainty is negligible compared to the reconstruction error in this case.  For a time-integrated analysis, the atmospheric background spatial dependency is assumed to be dependent on declination $\delta_i$ but uniform over right ascension. The product of the spatial component and the energy dependent background PDF $\epsilon_B$ at declination $\delta_i$ gives the background combined PDF $ \mathcal{B}_i(\vec{x_i},E_i,\delta_i) =  \mathcal{B}_{i}(\vec{x_i}) \epsilon_B(E_i,\delta_i)$. The background is constructed from the data itself.

Using these PDFs the likelihood is evaluated over all events in the declination band. If $n_S$ denotes the number of signal events in the declination band, the likelihood is:
\begin{equation}
    \mathcal{L}(\vec{x_s},n_S,\gamma) = \prod_i^N\left( \frac{n_S}{N}\mathcal{S}_i+(1-\frac{n_S}{N})\mathcal{B}_i \right) \,
\end{equation}
and the test statistic,where the $\hat{}$ notation is used to denote a best-fit ,is
\begin{equation}
   TS = -2 \, {\rm sign}(\hat{n_S}) \log\left[ \frac{\mathcal{L}(\vec{x_s},0)}{\mathcal{L}(\vec{x_s},\hat{n}_S,\hat{\gamma})} \right] \, .
\end{equation}

Our stacking study assumes that the time-integrated neutrino flux is directly proportional to the radio flux-densities observed at 15\, GHz. This is implemented by using the time-integrated radio observations (see distribution in Fig.~\ref{fig:aitoff}; right) as the weights for the stacking thereby changing the signal PDF to $\mathcal{S}_i= \frac{\sum_j \omega_j\mathcal{R}_j(\delta_j,\gamma) \mathcal{S}^j_i(\vec{x_s})}{\sum_j \omega_j\mathcal{R}_j(\delta_j,\gamma)}$, where $\omega_j$ is the weight given by the time-integrated flux-density of the $j^{\mathrm{th}}$ source and $\mathcal{R}_j(\delta_j,\gamma)$ is the detector weight at a source at declination of $\delta_j$ emitting neutrinos from a differential $E^{-\gamma}$ spectrum.

\begin{figure}
\centering
\includegraphics[width=.9\textwidth]{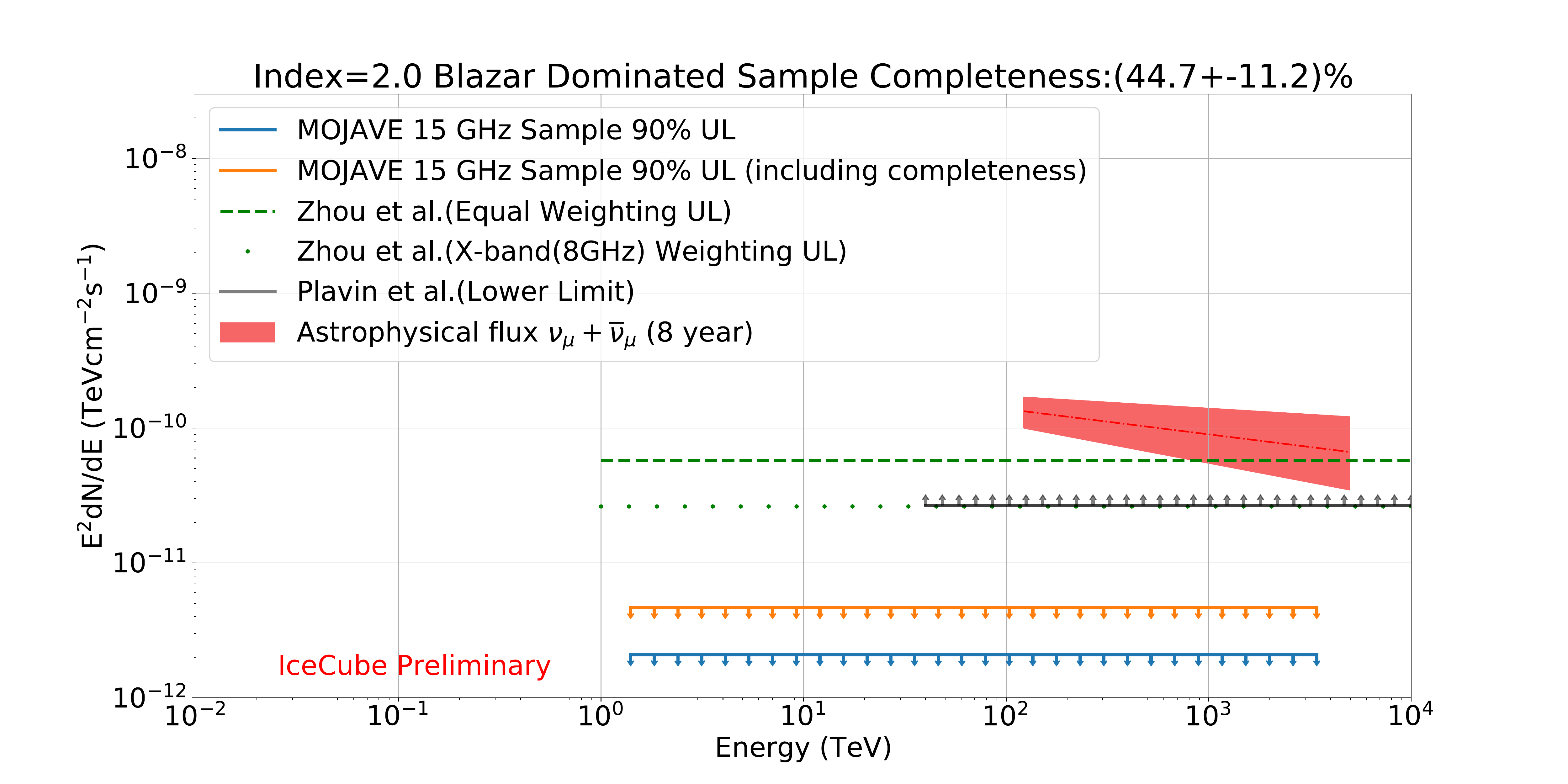}
\caption{The upper limits derived for the MOJAVE time integrated stacking work (by using 437 sources) for a spectral index of 2.0 are shown here. The energy range covered by our analysis (blue and orange) is derived using the region where 90\% of detected signal neutrinos would fall, under the assumption of an $E^{-2.0}$ spectrum. The upper limits shown by the green colored lines cover the energy range shown in Fig. 4 in \cite{zhou2021neutrino_agn} and the energy range covered by grey line is obtained using the 40\,TeV energy limit used by \cite{Plavin_2020b} to calculate the lower limit flux. Note that all three analyses (this work, \cite{zhou2021neutrino_agn} and \cite{Plavin_2020a,Plavin_2020b}) make use of different methods with different weighting schemes and source catalogs to study the correlation between radio observations and neutrinos making it difficult to make a direct comparison.
\vspace{-4.5mm}}
\label{fig:ul_2.0_blazar}
\end{figure}
\section{Results}

\begin{figure}
\centering
\includegraphics[width=.9\textwidth]{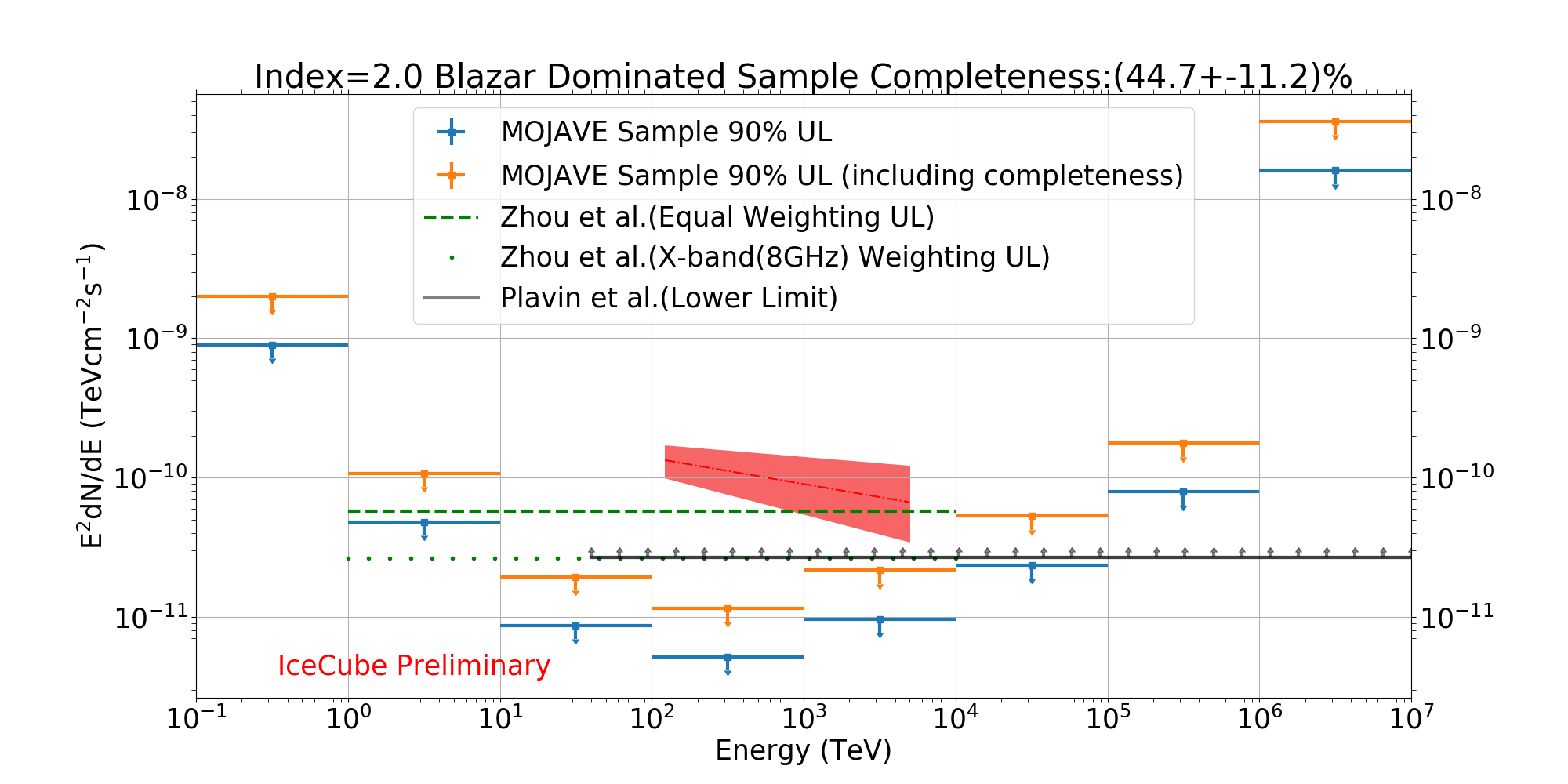}
\caption{The plot shows the 90\% differential upper limits derived by stacking the MOJAVE sources using the average radio flux as the weights for each energy bin. Note that all three analyses (this work, \cite{zhou2021neutrino_agn} and \cite{Plavin_2020a,Plavin_2020b}) make use of different methods with different weighting schemes and source catalogs to study the correlation between radio observations and neutrinos making it difficult to make a direct comparison.\vspace{-4.5mm}}
\label{fig:ul_2.0_diff}
\end{figure}

No significant evidence for a neutrino signal above the background expectation was seen for the time-integrated stacking analysis. While the contour plot in Fig.~\ref{fig:2dlikelihood} (right) shows the best-fit number of signal events ($n_s$) and spectral index ($\gamma$), the p-value that we obtained was 0.49. Because of the lack of a signal, we derive the upper limit on the muon neutrino flux with a 90\% confidence limit and show it alongside the recent results from \cite{zhou2021neutrino_agn,Plavin_2020a,Plavin_2020b} in Fig.~\ref{fig:ul_2.0_blazar} and Fig.~\ref{fig:ul_2.0_diff} using an unbroken power-law with spectral index=2.0. These results are shown alongside the diffuse muon neutrino flux reported by \cite{diffuse_flux_2017icecube}.

\vspace{1.5mm}
Out of the two theories for neutrino production for AGN ($pp$ and $p\gamma$), the $p\gamma$ neutrino production mechanism is favoured, since this analysis makes use of VLBA data which observes the region close to the core of the AGN. This is because, the radio emission from the parsec-scale jets is due to electron synchrotron emission. These synchrotron photons then undergo synchrotron self-Compton scattering to form the target photons that are seen in keV-MeV  and which undergo $p\gamma$ interactions to produce observable neutrinos. The neutrino spectrum will be affected by the target photon spectrum and will cut off at lower and higher energies because of either less interactions or lack of high-energy photons or pion cooling (see e.g. \cite{Kalashev2015,Plavin_2020a} for more details). Thus, it is important to also see how the neutrino flux behaves in different energy bins, which is shown in Fig.~\ref{fig:ul_2.0_diff}. For energies above 1\,PeV, more neutrinos undergo earth absorption which can be seen from the figure.

Recently \cite{zhou2021neutrino_agn} performed a study making use of 8\, GHz observations of radio-loud AGN found in the Radio Fundamental Catalog (RFC)\footnote{http://astrogeo.org/rfc/} to perform a stacking analysis similar to the one described here but using different weights. They provide an upper limit at 95\% C.L. for the stacking analysis because of a lack of significant results (denoted by green data points and dashed lines in Fig.~\ref{fig:ul_2.0_blazar} and ~\ref{fig:ul_2.0_diff}). While \cite{zhou2021neutrino_agn} use 8\, GHz measurements as the weights to check for a proportional correlation between 8\, GHz flux densities and neutrinos, we make use of the 15\, GHz observations to derive an upper limit at 90\% C.L. Both \cite{zhou2021neutrino_agn} and \cite{Plavin_2020a,Plavin_2020b} do not have the same constraints of the MOJAVE sample as this work, namely the source population being limited to radio loud AGN with bright cores. Moreover, because of different weighting schemes and data used, a detailed comparison between the three analysis is difficult.

The diffuse neutrino flux observed by the IceCube collaboration \cite{diffuse_flux_2017icecube} and shown in Fig.~\ref{fig:ul_2.0_blazar} and  ~\ref{fig:ul_2.0_diff} displays a unbroken power-law with spectral index of $\gamma$= 2.19. Similar to the procedure described by \cite{3fhl_icecube}, we calculate the maximum contribution of this blazar dominated radio sample to the diffuse flux under the assumption of a spectral index of 2.0. The energy range for our upper limits reflects the region where 90\% of detected signal neutrinos would fall. If, instead, we calculate the range where the analysis is most sensitive – by calculating threshold energies that degrade our sensitivity by 5\% – then this energy range shifts towards a higher energy range of 8\,GeV to 50\,PeV. By using the energy limits of the diffuse flux we find that the blazar dominated radio loud AGN sample, including completeness, can not explain more than $\sim 6.02\%$ of the  diffuse flux assuming a proportional correlation between the neutrino fluxes and the flux densities observed at 15\,GHz.

While our analysis fails to detect a  significant signal (similar to \cite{zhou2021neutrino_agn}), it is important to continue the search for correlations (or a lack of it) between radio observations of AGN and neutrino data observed with IceCube. This can be done by analyses of individual sources (see for example \cite{Maggi:2017e5} and \cite{OVRO_neutrinos_2020}) or a stacked search similar to this work. While a time-dependent analysis making use of the same methodology described above but using lightcurves from MOJAVE will be discussed in a separate work soon, the limitations of the MOJAVE dataset imply that observations from other radio telescopes like the Owens Valley Radio Observatory \cite{Richards_2011} or F-GAMMA program \cite{Angelakis_2019} will have to be used to get a better understanding in the future (see also \cite{Plavin_2020a}). 


\bibliographystyle{ICRC}
\bibliography{references}

\providecommand{\href}[2]{#2}\begingroup\raggedright\begin{thebibliography}{10}

\bibitem{icecube_2017_intro}
{\bfseries IceCube} Collaboration, M.~G. Aartsen {\em et~al.}
  \href{http://dx.doi.org/10.1088/1748-0221/12/03/P03012}{{\em JINST}
  {\bfseries 12} no.~03, (2017) P03012}.

\bibitem{icecube_2013_astrophysical_neutrino}
{\bfseries IceCube} Collaboration, M.~G. Aartsen {\em et~al.}
  \href{http://dx.doi.org/10.1126/science.1242856}{{\em Science} {\bfseries
  342} (2013) 1242856}.

\bibitem{TXS_Icecube}
{\bfseries IceCube, Fermi-LAT, MAGIC, AGILE, ASAS-SN, HAWC, H.E.S.S., INTEGRAL,
  Kanata, Kiso, Kapteyn, Liverpool Telescope, Subaru, Swift NuSTAR, VERITAS,
  VLA/17B-403} Collaboration, M.~G. Aartsen {\em et~al.}
  \href{http://dx.doi.org/10.1126/science.aat1378}{{\em Science} {\bfseries
  361} no.~6398, (2018) eaat1378}.

\bibitem{Plavin_2020a}
A.~Plavin, Y.~Y. Kovalev, Y.~A. Kovalev, and S.~Troitsky
  \href{http://dx.doi.org/10.3847/1538-4357/ab86bd}{{\em Astrophys. J.}
  {\bfseries 894} no.~2, (2020) 101}.

\bibitem{Plavin_2020b}
A.~V. Plavin, Y.~Y. Kovalev, Y.~A. Kovalev, and S.~V. Troitsky
  \href{http://dx.doi.org/10.3847/1538-4357/abceb8}{{\em Astrophys. J.}
  {\bfseries 908} no.~2, (2021) 157}.

\bibitem{OVRO_neutrinos_2020}
T.~Hovatta {\em et~al.}
  \href{http://dx.doi.org/10.1051/0004-6361/202039481}{{\em Astron. Astrophys.}
  {\bfseries 650} (2021) A83}.

\bibitem{zhou2021neutrino_agn}
B.~Zhou, M.~Kamionkowski, and Y.-f. Liang
  \href{http://dx.doi.org/10.1103/PhysRevD.103.123018}{{\em Phys. Rev. D}
  {\bfseries 103} no.~12, (2021) 123018}.

\bibitem{Eichler78}
D.~Eichler and D.~N. Schramm \href{http://dx.doi.org/10.1038/275704a0}{{\em
  Nature} {\bfseries 275} no.~5682, (1978) 704--706}.

\bibitem{pp_suppression_1987_sikora}
M.~{Sikora}, J.~G. {Kirk}, M.~C. {Begelman}, and P.~{Schneider}
  \href{http://dx.doi.org/10.1086/184980}{{\em Astrophys. J.} {\bfseries 320}
  (Sept., 1987) L81}.

\bibitem{Inoue_2019}
Y.~{Inoue}, D.~{Khangulyan}, S.~{Inoue}, and A.~{Doi}
  \href{http://dx.doi.org/10.3847/1538-4357/ab2715}{{\em Astrophys. J.}
  {\bfseries 880} no.~1, (July, 2019) 40}.

\bibitem{MOJAVEXV}
M.~L. Lister, M.~F. Aller, H.~D. Aller, M.~A. Hodge, D.~C. Homan, Y.~Y.
  Kovalev, A.~B. Pushkarev, and T.~Savolainen
  \href{http://dx.doi.org/10.3847/1538-4365/aa9c44}{{\em Astrophys. J. Suppl.}
  {\bfseries 234} no.~1, (Jan, 2018) 12}.

\bibitem{MOJAVEXVII}
M.~L. Lister, D.~C. Homan, T.~Hovatta, K.~I. Kellermann, S.~Kiehlmann, Y.~Y.
  Kovalev, W.~Max-Moerbeck, A.~B. Pushkarev, A.~C.~S. Readhead, and E.~Ros
  \href{http://dx.doi.org/10.3847/1538-4357/ab08ee}{{\em Astrophys. J.}
  {\bfseries 874} no.~1, (Mar, 2019) 43}.

\bibitem{Braun_2008}
J.~Braun, J.~Dumm, F.~De~Palma, C.~Finley, A.~Karle, and T.~Montaruli
  \href{http://dx.doi.org/10.1016/j.astropartphys.2008.02.007}{{\em Astropart.
  Phys.} {\bfseries 29} (2008) 299--305}.

\bibitem{diffuse_flux_2017icecube}
{\bfseries IceCube} Collaboration, M.~G. Aartsen {\em et~al.}, ``{The IceCube
  Neutrino Observatory - Contributions to ICRC 2017 Part II: Properties of the
  Atmospheric and Astrophysical Neutrino Flux},''
\newblock 10, 2017.
\newblock \href{http://arxiv.org/abs/1710.01191}{{\ttfamily arXiv:1710.01191
  [astro-ph.HE]}}.

\bibitem{Kalashev2015}
O.~Kalashev, D.~Semikoz, and I.~Tkachev
  \href{http://dx.doi.org/10.1134/S106377611503022X}{{\em J. Exp. Theor. Phys.}
  {\bfseries 120} no.~3, (2015) 541--548}.

\bibitem{3fhl_icecube}
{\bfseries IceCube} Collaboration, M.~Huber
  \href{http://dx.doi.org/10.22323/1.358.0916}{{\em PoS} {\bfseries ICRC2019}
  (2020) 916}.

\bibitem{Maggi:2017e5}
{\bfseries IceCube} Collaboration, G.~Maggi, K.~D. de~Vries, and N.~van
  Eijndhoven \href{http://dx.doi.org/10.22323/1.301.1000}{{\em PoS} {\bfseries
  ICRC2017} (2018) 1000}.

\bibitem{Richards_2011}
J.~L. Richards {\em et~al.}
  \href{http://dx.doi.org/10.1088/0067-0049/194/2/29}{{\em Astrophys. J.
  Suppl.} {\bfseries 194} (2011) 29}.

\bibitem{Angelakis_2019}
E.~Angelakis, L.~Fuhrmann, I.~Myserlis, J.~A. Zensus, I.~Nestoras,
  V.~Karamanavis, N.~Marchili, T.~P. Krichbaum, A.~Kraus, and J.~P. Rachen
  \href{http://dx.doi.org/10.1051/0004-6361/201834363}{{\em Astron. Astrophys.}
  {\bfseries 626} (Jun, 2019) A60}.

\end{thebibliography}\endgroup

\clearpage
\section*{Full Author List: IceCube Collaboration}




\scriptsize
\noindent
R. Abbasi$^{17}$,
M. Ackermann$^{59}$,
J. Adams$^{18}$,
J. A. Aguilar$^{12}$,
M. Ahlers$^{22}$,
M. Ahrens$^{50}$,
C. Alispach$^{28}$,
A. A. Alves Jr.$^{31}$,
N. M. Amin$^{42}$,
R. An$^{14}$
% ...
\\

\noindent
% ...
$^{12}$ Universit{\'e} Libre de Bruxelles, Science Faculty CP230, B-1050 Brussels, Belgium \\
% ...
$^{14}$ Department of Physics and Laboratory for Particle Physics and Cosmology, Harvard University, Cambridge, MA 02138, USA \\
% ...
$^{17}$ Department of Physics, Loyola University Chicago, Chicago, IL 60660, USA \\
$^{18}$ Dept. of Physics and Astronomy, University of Canterbury, Private Bag 4800, Christchurch, New Zealand \\
% ...
$^{22}$ Niels Bohr Institute, University of Copenhagen, DK-2100 Copenhagen, Denmark \\
% ...
$^{28}$ D{\'e}partement de physique nucl{\'e}aire et corpusculaire, Universit{\'e} de Gen{\`e}ve, CH-1211 Gen{\`e}ve, Switzerland \\
% ...
$^{31}$ Karlsruhe Institute of Technology, Institute for Astroparticle Physics, D-76021 Karlsruhe, Germany  \\
% ...
$^{42}$ Bartol Research Institute and Dept. of Physics and Astronomy, University of Delaware, Newark, DE 19716, USA \\
% ...
$^{50}$ Oskar Klein Centre and Dept. of Physics, Stockholm University, SE-10691 Stockholm, Sweden \\
% ...
$^{59}$ DESY, D-15738 Zeuthen, Germany \\
% ...

\subsection*{Acknowledgements}

\noindent
USA {\textendash} U.S. National Science Foundation-Office of Polar Programs,
U.S. National Science Foundation-Physics Division,
U.S. National Science Foundation-EPSCoR,
Wisconsin Alumni Research Foundation,
Center for High Throughput Computing (CHTC) at the University of Wisconsin{\textendash}Madison,
Open Science Grid (OSG),
Extreme Science and Engineering Discovery Environment (XSEDE),
Frontera computing project at the Texas Advanced Computing Center,
U.S. Department of Energy-National Energy Research Scientific Computing Center,
Particle astrophysics research computing center at the University of Maryland,
Institute for Cyber-Enabled Research at Michigan State University,
and Astroparticle physics computational facility at Marquette University;
Belgium {\textendash} Funds for Scientific Research (FRS-FNRS and FWO),
FWO Odysseus and Big Science programmes,
and Belgian Federal Science Policy Office (Belspo);
Germany {\textendash} Bundesministerium f{\"u}r Bildung und Forschung (BMBF),
Deutsche Forschungsgemeinschaft (DFG),
Helmholtz Alliance for Astroparticle Physics (HAP),
Initiative and Networking Fund of the Helmholtz Association,
Deutsches Elektronen Synchrotron (DESY),
and High Performance Computing cluster of the RWTH Aachen;
Sweden {\textendash} Swedish Research Council,
Swedish Polar Research Secretariat,
Swedish National Infrastructure for Computing (SNIC),
and Knut and Alice Wallenberg Foundation;
Australia {\textendash} Australian Research Council;
Canada {\textendash} Natural Sciences and Engineering Research Council of Canada,
Calcul Qu{\'e}bec, Compute Ontario, Canada Foundation for Innovation, WestGrid, and Compute Canada;
Denmark {\textendash} Villum Fonden and Carlsberg Foundation;
New Zealand {\textendash} Marsden Fund;
Japan {\textendash} Japan Society for Promotion of Science (JSPS)
and Institute for Global Prominent Research (IGPR) of Chiba University;
Korea {\textendash} National Research Foundation of Korea (NRF);
Switzerland {\textendash} Swiss National Science Foundation (SNSF);
United Kingdom {\textendash} Department of Physics, University of Oxford.

\end{document}